\documentclass[pdfusetitle,aps,prl,twocolumn,showpacs,nofootinbib,preprintnumbers,superscriptaddress,groupedaddress]{revtex4-1}
\pdfoutput=1
 \usepackage{hyperref}
\hypersetup{colorlinks=true,linkcolor=magenta,anchorcolor=green,citecolor=cyan,filecolor=black,menucolor=black,urlcolor=black}
\usepackage{stackrel}
\usepackage[compat=1.1.0]{tikz-feynman}
\usepackage{tikz,contour}
\usetikzlibrary{calc,decorations.markings}

\usepackage[force]{feynmp-auto}

\usepackage{bbm}
\usepackage{bbold}
\usepackage{float}

\definecolor{linkblue}{rgb}{0.1,0.3,.7}
\definecolor{forestgreen(web)}{rgb}{0.13, 0.55, 0.13}
\definecolor{lava}{rgb}{0.81, 0.06, 0.13}
\hypersetup{ 
breaklinks,
colorlinks,
citecolor=forestgreen(web),
filecolor=linkblue,
linkcolor=lava,
urlcolor=linkblue
}

\usepackage{caption}
\usepackage{subcaption}
\usepackage{simpler-wick}

\usepackage{pgfplots}
\usepackage[T1]{fontenc}
\usepackage[]{slashed}
\usepackage[]{bm}     
\usepackage{physics}
\usepackage{lipsum}
\usepackage{dsfont}
\usepackage{soul}

\usepackage{bbm,amsmath,graphicx,amssymb,amsfonts,amsthm}

\definecolor{bubbles}{rgb}{0.91, 1.0, 1.0}
\definecolor{aquamarine}{rgb}{0.5, 1.0, 0.83}
\definecolor{bubblegum}{rgb}{0.99, 0.76, 0.8}
\definecolor{blackbell}{rgb}{0.64, 0.64, 0.82}
\definecolor{dollarbill}{rgb}{0.72, 0.93, 0.6}

\pgfdeclarelayer{bg}    
\pgfsetlayers{bg,main}  

\definecolor{Mathematica}{HTML}{ed192d}

\usepackage{float}
\usepackage[T1]{fontenc} 

\begin{document}
\title{On Duality Invariant Yang-Mills Theory}

\author{Carlo Alberto Cremonini, Erik Hundeshagen and Ivo Sachs}
\affiliation{Arnold-Sommerfeld-Center for Theoretical Physics, Ludwig-Maximilians-Universit\"at M\"unchen,
        Theresienstr. 37, D-80333 Munich, Germany}

\date{\today}

\begin{abstract} 
We provide an explicit construction of a manifestly duality invariant, interacting deformation of Maxwell theory in four dimensions in terms of mutually local, but interacting 1- and 3-forms. Interestingly, our theory is formulated directly as a BRST quantized gauge theory, while the underlying gauge invariant Lagrangian before gauge fixing is obscured. Furthermore, the underlying gauge invariance is based on an associative, rather than a Lie symmetry.
\end{abstract}
\pacs{}

\maketitle

\section{Introduction}

Since Maxwell's equations are invariant under the exchange of electric and magnetic fields, it is natural to speculate whether electric-magnetic duality can be extended to an interacting theory. For a single Maxwell field, it is long known to be the case. Schrödinger noticed that the Born-Infeld action has this property. See \cite{Gibbons:1995cv} for an extensive discussion on this topic. On the other hand, Yang-Mills theory does not have a local and duality-invariant formulation. See e.g. \cite{Bekaert:2001wa,Sorokin:2025ezs} and references therein. 
Below, we will present a manifestly duality invariant non-abelian interacting deformation of Maxwell theory, involving a finite number of auxiliary fields. By doing so, we avoid the no-go theorem in \cite{Bekaert:2001wa} in two aspects. First, the interacting theory is not a local functional of one vector potential. It is, however, a local functional of a 1-form and a 3-form potential combined with an auxiliary field. Second, while the free theory is manifestly invariant under Maxwell gauge transformations, the gauge invariance is not manifest in the interacting theory. Rather, the interacting theory is described as a deformation of the gauge fixed BRST-fication of the free theory. In other words, the deformation is formulated directly in BRST-quantization rather than a gauge theory, which is later BRST quantized. It is not clear to us whether our model has a formulation as a non-abelian gauge theory in the usual sense. 

A key new ingredient in our approach, which was not present in previous literature, is that we replace the wedge product by a Clifford product of forms and the exterior differential $\mathrm{d}$ is replaced by the K\"ahler-Dirac operator $K=\mathrm{d}+\mathrm{d}^\dagger$. In a way, $K$ is the natural choice since it makes Hodge duality for the kinetic term manifest. However, to include interactions, we need a multiplication that is compatible with Hodge duality.  The Clifford product is obtained simply by substituting differentials $\mathrm{d}x^\mu$ by a gamma matrix $\gamma^\mu$. In fact, this product turns out not to be invariant under Hodge duality, but its projection on top-forms (i.e. under the integral) is. We will find that this is sufficient to obtain a manifestly electric-magnetic invariant form of the action. An immediate consequence of introducing the Clifford product is that it induces not only Lie brackets of the gauge-generators, but also anti-commutators as well. This means that admissible gauge Lie algebras need to admit an associative structure as well. Such is the case for the gauge group $U(n)^{\mathbb{C}}$, for example. Another option is to consider supergroups instead of ordinary gauge groups.

\section{Free Theory}
Let us begin with the familiar Maxwell action (the notations adopted throughout the text are explained in the last section, and in the whole text, the trace over the Lie algebra generators is understood, when applicable)
\begin{align}
  S_1= \frac{1}{2} \int (\mathrm{d}A^{[1]}, \mathrm{d}A^{[1]})\ ,
\end{align}
and the "dual" formulation in terms of a 3-form 
\begin{align}
  S_2=  \frac{1}{2}\int (\mathrm{d}^\dagger A^{[3]}, \mathrm{d}^\dagger A^{[3]}) \ ,
\end{align}
where $\int(\cdot,\cdot)$ is the Hodge inner product.
Here we mean dual in the sense that if we write $A^{[3]}=*\tilde A^{[1]}$ and interpret $\tilde A^{[1]}$ as the potential naturally coupling to magnetic 4-current, then $S_2$ naturally describes the dynamics of an electromagnetic field coupled to magnetic charges.

To continue, we introduce an auxiliary 2-form to couple the two field strengths $\mathrm{d}A^{[1]}$ and $\mathrm{d}^\dagger A^{[3]}$. Concretely, we set 
\begin{align}\label{eq:Slin}
    S=S_1+S_2+i\int& (\mathrm{d} A^{[1]} + \mathrm{d}^\dagger A^{[3]} , B^{[2]} ) \ .
\end{align}
Variation w.r.t. $B^{[2]}$ implies 
\begin{align}\label{eq:pairingthelinearfieldstrengths}
  \mathrm{d} A^{[1]}+\mathrm{d}^\dagger A^{[3]}=0\ ,  
\end{align}
thus identifying the two field strengths $\mathrm{d}A^{[1]}$ and $\mathrm{d}^\dagger A^{[3]}$.  
Note that this is the generic duality symmetric coupling at linear order in the absence of boundaries. Before proceeding with the analysis of the equations of motion, let us consider the naive counting of the theory described by two Maxwell theories with the constraint identifying the two field-strengths. For $A^{[3]}=*\tilde A^{[1]}$ one recovers a duality invariant action as in \cite{Schwarz:1993vs} and \cite{Pasti:1995ii} for instance  however, with an extra auxiliary field. The coupling \eqref{eq:pairingthelinearfieldstrengths} is implied in the quantization of the relativistic spinning particle \cite{Cremonini:2025eds}. In string theory, such a two-form arises as a massive field in the spectrum.

We then expect that this identifies the degrees of freedom of the two Maxwell systems, such that of the 4+4 components of $A^{[1]}$ and $A^{[3]}$, only 4 are independent. The gauge symmetry of the theory then allows us to identify two propagating degrees of freedom, as usual in Maxwell theory. If, furthermore, $\tilde A^{[1]}= A^{[1]}$, we recover the theory of a chiral 2-form, e.g. \cite{Bengtsson:1996fm} with no local degrees of freedom. We should note that the variation w.r.t. $A$, on shell, furthermore implies $\mathrm{d}B^{[2]}=\mathrm{d}^{\dagger} B^{[2]}=0$ which suggests that $B^{[2]}$ is the field strength of another Maxwell field. However, the components of $B^{[2]}$ can be gauged away thanks to an on-shell 2-form symmetry, analogous to the residual gauge in Maxwell after imposing the Lorentz gauge-fixing condition. We could then include ghosts (and ghosts for ghosts as in \cite{Bengtsson:1996fm}) for these modes, but since $B^{[2]}$ decouples, as we will see below, this won't be necessary. 

To see that this intuition is correct, let us analyze the degrees of freedom in terms of constraints in phase space (see also \cite{Bengtsson:1996fm}).
From \eqref{eq:Slin}, we read the momenta as
\begin{align}
    \pi^0= 0 \ , \ \pi^i = \partial^i A^0 - \partial^0 A^i + 2 B^{0i} \ , \ \pi^{\mu \nu} = 0 \ , \ \ldots
\end{align}
where $\pi^{\mu \nu}$ are the six momenta conjugate to $B_{\mu \nu}$ and "$\ldots$" denotes analogous expressions for the momenta of the other Maxwell system, which will be denoted by $\tilde \pi$. The introduction of the Lagrange multiplier $B^{[2]}$ in the action leads to six more primary constraints, compared to the usual analysis for the two Maxwell theories with $\phi_1 = \pi^0, \tilde \phi_1 = \tilde \pi^0$, given by the momenta $\phi_{\mu \nu} = \pi_{\mu \nu}$. The Hamiltonian on the constraint surface reads 
\begin{align}
    \nonumber H_0 = & \frac{1}{2} \pi^i \pi_i + \frac{1}{2} \pi^i \partial_i A_0 - \frac{1}{4} F_{ij}F^{ij} + \\
    & - 6 B_{0i} B^{0i} - B_{ij} \left( F^{ij} + \tilde{F}^{ij} \right) + \ldots \ ,
\end{align}
and it is extended as
\begin{align}
    H_E = H_0 + g_1 \phi_1 + \tilde{g}_1 \tilde{\phi}_1 + g^{\mu \nu} \phi_{\mu \nu} \ .
\end{align}
These eight primary constraints generate, by consistency, eight more secondary constraints as
\begin{subequations}
\begin{align}
    \left\lbrace H_E , \pi^0 \right\rbrace = - \partial_i \pi^i \to \partial_i \pi^i \equiv \phi_{2} \ , \\
    \left\lbrace H_E , \tilde \pi^0 \right\rbrace = - \partial_i \tilde \pi^i \to \partial_i \tilde \pi^i \equiv \tilde \phi_{2} \ , \\
    \left\lbrace H_E , \pi_{0i} \right\rbrace = -6 B_{0i} \to B_{0i} \equiv \phi_{sec, i} \ , \\
    \left\lbrace H_E , \pi_{ij} \right\rbrace = - \left( F_{ij} + \tilde F_{ij} \right) \to F_{ij} + \tilde F_{ij} \equiv \phi_{sec, ij} \ .
\end{align}
\end{subequations}
The Hamiltonian is thus extended as
\begin{align}
    H_{E}' = H_E + g_2 \phi_2 + \tilde g_2 \tilde \phi_2 + g_{0i} \phi_{sec,i} + g_{ij} \phi_{sec,ij} \ .
\end{align}
First of all, notice that $\phi_{sec,i}$ do not generate other constraints as
\begin{align}
    \left\lbrace H_E' , \phi_{sec,i} \right\rbrace = g_{0i} \approx 0 \ .
\end{align}
On the other hand, $\phi_{sec,ij}$ generate more constraints as
\begin{align}
    \left\lbrace H_E' , \phi_{sec,ij} \right\rbrace  \approx - \partial_i \left( \pi_j + \tilde \pi_j \right) + \partial_j \left( \pi_i + \tilde \pi_i \right) \equiv \phi_{ter,ij} \ ,
\end{align}
and $\left\lbrace H_E' , \phi_{ter,ij} \right\rbrace  \approx 0$. It is not hard to see that $\phi_1, \tilde \phi_1 , \phi_2, \tilde \phi_2$ and $\phi_{ij}$ are first-class constraints, while $\phi_{0i}, \phi_{sec,i}, \phi_{sec,ij} $ and $\phi_{ter,ij}$ are second-class constraints. One subtlety in counting the number of constraints is that the combination $\phi_2 + \tilde \phi_2$, $\phi_{sec,ij}$ and $\phi_{ter,ij}$ contribute altogether to reduce the number of canonical variables by six, instead of eight, as one could think by counting them as a first-class constraint and six second-class constraint. However, a more detailed analysis, shows that these constraints lead to the identification of the electric and magnetic fields of the two theories, that is, they fix six canonical variables\footnote{This can be analogously seen by turning off the second Maxwell system and thus considering in the action the remaining term $B_{\mu \nu} F^{\mu \nu}$. In this case, we obtain that both the electric and magnetic fields are zero on the constraint surface.}. We conclude this analysis with a final remark on the interacting theory, which will be explored in the next section. In the previous analysis, we verified that the introduction of the constraint with a two-form Lagrange multiplier does not introduce new degrees of freedom. On the other hand, in the interacting theory, the presence of $B$ results in a cubic vertex with the structure $BAA$. However, the interacting equations of motion allow $B=0$ as a consistent solution, thus suggesting that even if $B$ were contributing with propagating degrees of freedom, these would decouple from the rest of the theory. This mechanism resembles the one studied in \cite{Sen:2015nph,Sen:2019qit}, with the difference that there is no self-duality constraint on $B$.

To continue, we introduce two more auxiliary fields as Lagrange multipliers enforcing gauge-fixing conditions as
\begin{align}\label{eq:gaugefixedactionnoghosts}
    S=S_1+S_2+i\int&(\mathrm{d}^\dagger A^{[1]},B^{[0]})+(\mathrm{d} A^{[1]}+\mathrm{d}^\dagger A^{[3]} ,B^{[2]})\nonumber\\
    &+ (\mathrm{d} A^{[3]},B^{[4]}) \ .
\end{align}
The equations of motion resulting from variation w.r.t. $A$ are then 
\begin{align}\label{eq:eml}
    \Box A^{[1]} + i \mathrm{d} B^{[0]} + i \mathrm{d}^\dagger B^{[2]} = 0 \ , \nonumber\\
    \Box A^{[3]} + i \mathrm{d} B^{[2]} + i \mathrm{d}^\dagger B^{[4]} = 0 \ .
\end{align}
Combining this with \eqref{eq:pairingthelinearfieldstrengths} $A^{[3]}$ is (non-locally) expressed in terms of $A^{[1]}$ or vice versa at the level of equations of motion. Finally, variation w.r.t. $B^{[0]}$ and $B^{[4]}$ implies
\begin{align}
    \mathrm{d}^\dagger A^{[1]} = 0 =\mathrm{d} A^{[3]} \ ,
\end{align}
thus enforcing the Lorentz gauge as an equation of motion. In order to complete the BRST quantization of \eqref{eq:gaugefixedactionnoghosts}, we need the \emph{ghost fields}, as for the usual abelian theory. The complete gauge-fixed action reads
\begin{align}\label{eq:fullgaugefixedaction}
    S=&S_1+S_2+i\int (\mathrm{d}^\dagger A^{[1]},B^{[0]})+(\mathrm{d} A^{[1]}+\mathrm{d}^\dagger A^{[3]} ,B^{[2]})\nonumber\\
    &+ (\mathrm{d} A^{[3]},B^{[4]}) + \left( b^{[0]} , \Box c^{[0]} \right) + \left( b^{[4]} , \Box c^{[4]} \right) \ .
\end{align}
Notice that the constraint analysis of the action in \eqref{eq:gaugefixedactionnoghosts} would lead to two extra degrees of freedom, which are exactly removed by the introduction of the two ghost systems.

The action in \eqref{eq:gaugefixedactionnoghosts}, or rather an equivalent one where the kinetic terms for the two Maxwells are written as $(A^{[i]} , \Box A^{[i]} )$, since we are  imposing the Lorentz gauge, can be written in compact form by means of the \emph{multiforms} $A=A^{[1]}+A^{[3]}$ and $B=B^{[0]}+B^{[2]}+B^{[4]}$, and of the K\"ahler-Dirac operator $K = d + d^\dagger$ as
\begin{align}\label{eq:actionB}
    S=\int \frac{1}{2}(A,K^2 A)+i(KA,B) \ .
\end{align}
The Hodge inner product is also defined for multiforms since the integral picks out top-forms which are precisely the contractions $(\cdot,\cdot)$ of components of the same degree.
The equations of motion then take the compact form 
\begin{align}
    K^2A +iKB = \square A+ iKB = 0, \quad KA = 0\ .
\end{align}
We can analogously recast \eqref{eq:actionB} in a more familiar form in terms of (a suitable extension of) the field strengths. Indeed, we can define
\begin{align}\label{eq:linearfieldstrengths}
    F = K A^{[1]} = F^{[0]} + F^{[2]} , \ \tilde{F} = K A^{[3]} = \tilde{F}^{[2]} + \tilde{F}^{[4]} \ ,
\end{align}
thus leading to
\begin{align}\label{eq:actionC}
    S=\int \frac{1}{2}(F + \tilde{F},F + \tilde{F})+i(F + \tilde{F},B) \ .
\end{align}
Note that we refer to \eqref{eq:linearfieldstrengths} as "field strengths" with an abuse of notation: they are indeed not gauge-invariant objects, as they include the Lorentz gauge in their structure. 
Despite the appearance of $(F+\tilde F)^2$ in \eqref{eq:actionC}, this term does not couple the two field strengths due to $\mathrm d^2 = (\mathrm{d}^\dagger)^2 = 0$, and only the $B$-field term relates the two.

\section{Interactions}

In this section, we show that the description in terms of multiforms and the K\"ahler-Dirac operator naturally encodes the extension to the interacting case. This is achieved by replacing the usual (wedge) product with the \emph{Clifford product}, which is, by construction, compatible with the duality we aim to achieve.

Let us first recall the Clifford product of forms: from an algorithmic point of view, we define it by the substitution 
\begin{align}
    A^{[n]}=A_{\mu_1\cdots \mu_n}\mathrm{d}x^{\mu_1}\wedge \cdots \wedge \mathrm{d}x^{\mu_1}\to A_{\mu_1\cdots \mu_n} \psi^{\mu_1}\cdots \psi^{\mu_n}
\end{align}
where $A_{\mu_1\cdots \mu_n}$ is totally antisymmetric in its indices, and then use Wick's theorem to contract the $\psi$ symbols in all possible combinations. This means that the Clifford product, denoted by "$\vee$", is defined as 
\begin{align}
    A^{[n]}\vee B^{[m]}=&
    A_{\mu_1\cdots \mu_n}B_{\mu_{n+1},\cdots \mu_{n+m}}\nonumber\\&\left(:\psi^{\mu_1}\cdots \psi^{\mu_n}\psi^{\mu_{n+1}}\cdots \psi^{\mu_{n+m}}: \right.\nonumber\\
    &\left. + \wick{:\psi^{\mu_1}\cdots\c1\psi^k \cdots\c1\psi^p }\cdots \psi^{\mu_{n+m}}:\right.\nonumber\\
    &\left.+\wick{:\psi^{\mu_1}\cdot\c1\psi^k \cdots\c2\psi^p\cdots\c1\psi^q \cdots\c2\psi^r}\cdot \psi^{\mu_{n+m}}:\right.\nonumber\\
    &\left.+\cdots\right)\,,
\end{align}
where $\wick{\c1\psi^k \c1\psi^p }=2\eta^{kp}$ and it is understood that normal ordered monomials $:\psi^{\mu_1}\cdots \psi^{\mu_p}:$ are totally antisymmetric in all indices. One thus substitutes back $\psi^i \mapsto dx^i$ to get a differential form. With this product, we can then introduce a \emph{covariant} Kähler-Dirac operator $K_A$ defined as 
\begin{align}\label{eq:nonabelianKD}
    K_A=K-iA\vee \ ,
\end{align}
where $A=A^{[1]}+A^{[3]}$ is a multiform as in the previous section. By the nature of $\vee$, $K_A$ is an operator mapping multiforms to multiforms. Note that $K$ does not satisfy Leibnitz's property w.r.t. to $\wedge$ or $\vee$. Crucially, $K_A$ requires no choice of $A^{[1]}$ over $A^{[3]}$ or vice-versa, thus we can use this operator to introduce interactions to our theory in a manifestly duality-symmetric fashion.

We can now define an interacting theory by simply replacing $K$ by $K_A$ in \eqref{eq:actionB} as
\begin{align}\label{eq:nlactionB}
    S=&\int \frac{1}{2}(A,(K_A)^2A)+i(K_AA,B)\nonumber\\
    =&\int \frac{1}{2}(K_AA,K_AA)+i(K_A A,B) \ ,
\end{align}
where the equality of the two expressions, up to boundary terms, follows from the associativity of $\vee$ and the cyclicity of the inner product. We can also rewrite the action in polynomial form as
\begin{align}
    S=\int &\frac{1}{2}(KA,KA)+i(K A,B) -\frac{i}{2}(A\vee A,(KA))\nonumber\\
    &-\frac{i}{2}(A,K(A\vee A))- \frac{1}{2}(A\vee A,A\vee A)+(A\vee A,B) \ .
\end{align}
Again, we can denote
\begin{align}
    F = K_A A^{[1]} \,, \quad \tilde F = K_A A^{[3]} \ ,
\end{align}
both of which now contain components in all even degrees. Then, we can again write the action \eqref{eq:nlactionB} in the compact form
\begin{equation}\label{eq:actionCnonlinear}
    S = \int\frac{1}{2} (F+\tilde F, F+\tilde F) +i(F+\tilde F, B).
\end{equation}
The equations of motion resulting from the variation w.r.t. $B$ are
\begin{align}
    K_A A = 0 \ .
\end{align}
More explicitly, they result in the non-linear equations given in \cite{Cremonini:2025eds}:
\begin{subequations}\label{eq:eomfromBnonlinear}
\begin{align}
  \mathrm{d}^\dagger A^{[1]} - i \left. A \vee A\right|_{0} = 0 \ ,\label{eq:eomfromBnonlineara} \\
     \mathrm{d} A^{[1]} + \mathrm{d}^\dagger A^{[3]} - i \left. A \vee A\right|_{2} = 0 \ , \label{eq:eomfromBnonlinearb} \\  
    \mathrm{d} A^{[3]} - i \left. A \vee A\right|_{4} = 0 \ , \label{eq:eomfromBnonlinearc}
\end{align}
\end{subequations}
where we denote the projection to the degree $p$ component of the multiform $X$ with $\left. X \right|_p$. Eqn's \eqref{eq:eomfromBnonlineara} and \eqref{eq:eomfromBnonlinearc} are to be interpreted as non-linear gauge fixing conditions, while \eqref{eq:eomfromBnonlinearb} generalizes \eqref{eq:pairingthelinearfieldstrengths}, relating $A^{[3]}$ and $A^{[1]}$. Altogteher, \eqref{eq:eomfromBnonlinear} encodes the Yand-Mills equations for the gauge field $A^{[1]}$ (resp., $A^{[3]}$) deformed by the non-local contributions coming from $A^{[3]}$ (resp., $A^{[1]}$), as can be seen by acting on the second equation in \eqref{eq:eomfromBnonlinear} with $\mathrm{d}$ (resp., $\mathrm{d}^\dagger$).

Notice that now there are terms in the action \eqref{eq:nlactionB} that mix $A^{[1]}$ and $A^{[3]}$ without the use of the $B$-field. However, this can be traced back to the fact that the covariant $K_A$ uses both $A^{[1]}$ and $A^{[3]}$. Therefore, the mixing terms appear since either gauge potential resembles a "background" for the other field strength, respectively. Hence, the structure from \eqref{eq:actionC} carries over to \eqref{eq:actionCnonlinear} despite the mixing.

Thus, the action \eqref{eq:nlactionB} realizes a gauge-invariant non-abelian action keeping the electro-magnetic duality between $A^{[1]}$ and $A^{[3]}$ manifest. This is compatible with the no-go theorem in \cite{Bekaert:2001wa} due to the (on-shell) non-local relation between the two gauge fields $A^{[1]}$ and $A^{[3]}$. However, as long as we keep the auxiliary $B$-field, the action is duality invariant without non-local interaction terms. This should be a good starting point for quantization, although we will not pursue this at present. 
 
\section{Gauge Structure}

Let us first reconsider the quadratic action \eqref{eq:fullgaugefixedaction}. As pointed out above, it represents a fully gauge-fixed action that can be expressed in terms of  multiforms as
\begin{align}\label{eq:abeliangfactionwithmultiforms}
    S=\int \frac{1}{2}(A,K^2 A)+i(KA,B) + \left( b , \Box c \right) \ ,
\end{align}
where $b = b^{[0]} + b^{[4]}$ and $c =  c^{[0]} + c^{[4]}$. Notice, in particular, that these multiforms do not include the 2-form components $b^{[2]}$ and $c^{[2]}$, as the constraint analysis shows that we need a single ghost (for each copy of the system) as in the usual analysis of Maxwell theory.
Next, we give a gauge-fixing fermion, which leads to our gauge-fixed action. We start with the BRST transformation for the two copies of Maxwell theory in a multiform description:
\begin{align}
    \delta_{BRST} A = sA = Kc,\quad sc = 0,
\end{align}
and introduce two trivial pairs
\begin{align}
    sb &= s(b^{[0]} + b^{[2]} + b^{[4]}) = i(B^{[0]} + B^{[2]} + B^{[4]}) = iB \, , \\
    s \tilde b &= s(\tilde b^{[0]} + b^{[4]}) = \lambda ^{[0]} + \lambda^{[4]} = \lambda \,, \\
    &s B = 0 \, , \, s \lambda = 0 \ .
\end{align}
We take the gauge-fixing fermion
\begin{align}\label{eq:gaugefixingfermion}
    \Psi = (b,KA) - (\tilde b,\frac 12\lambda + KA)\ ,
\end{align}
the components of which enforce both the constraint between the two Maxwell theories as well as Lorenz gauge for both Maxwell copies. We get
\begin{align}
    s\Psi = i(B,KA) - \frac 12(\lambda,\lambda) - (\lambda, KA) + (b-\tilde b, K^2 c)\ ,
\end{align}
which after integrating out $\lambda^{[0]}$ and $\lambda^{[4]}$ becomes
\begin{align}
    s\Psi &= i(B,KA) + \frac12(\mathrm d^\dagger A^{[1]},\mathrm d^\dagger A^{[1]}) + \\
    &+\frac12 (\mathrm d A^{[3]}, \mathrm d A^{[3]}) + (b - \tilde b, K^2 c) \ .
\end{align}
Adding this to the two copies of Maxwell yields precisely the gauge-fixed and constraint action
\begin{align}
    S &= \int (\mathrm d A^{[1]}, \mathrm d A^{[1]}) + (\mathrm d^\dagger A^{[3]}, \mathrm d^\dagger A^{[3]}) + s\Psi \nonumber\\
    &= \int \frac12 (A,K^2A) + (B,KA) +  (b -\tilde b, K^2 c)\ .
\end{align}
We can absorb $\tilde b$ into $b$ to arrive at \eqref{eq:abeliangfactionwithmultiforms}. Because we integrated out the auxiliary $\lambda$, the BRST transformations that leave this action invariant are now
\begin{align}
    sA &= Kc, & sc &= 0, & sb &= iB + \mathrm d^\dagger A^{[1]} + \mathrm d A^{[3]}, & s B &= 0 \ .
\end{align}
Note that while $b$ contains a two-form component $b^{[2]}$, it only appears through a boundary term  $\left( d b^{[2]} , d^\dagger c^{[4]} \right) + \left( d^\dagger b^{[2]} , d c^{[0]} \right)$.

The extension to the non-linear theory now follows as in the previous section: we start from the action (the trace over the gauge algebra is understood) 
\begin{align}
    S=\int \frac{1}{2}(A,K_A^2 A) + \left( A^* , K_A c \right) + \left( b^*,B \right) \ ,
\end{align}
where we lifted the K\"ahler-Dirac operator as in \eqref{eq:nonabelianKD}
. The extended gauge-fixing fermion can be chosen as
\begin{align}
    \Psi = b K_A A \ .
\end{align}
Again, this imposes the constraint on the two field strengths as well as a non-linear modification of the Lorentz gauge-fixing condition \eqref{eq:eomfromBnonlineara} and \eqref{eq:eomfromBnonlinearc}. These non-standard conditions match the results obtained from the geometric analysis of the supermoduli space of the spinning particle considered in \cite{Cremonini:2025eds}.

\begin{acknowledgments}{\bf Acknowledgments.}
We would like to thank Dimitri Sorokin for sharing numerous insights on a first draft of this paper. This work is supported by the Excellence Cluster Origins of the DFG under Germany’s Excellence Strategy EXC-2094 390783311 as well as EXC 2094/2: ORIGINS 2.
\end{acknowledgments}

\appendix
\section{Notation}
In this section, we show two properties of the Hodge inner product for multiforms, which we have used to simplify the action for the interacting theory, namely its cyclicity with the Clifford product and that we can integrate the Kähler-Dirac operator by parts.

We have used the Hodge inner product, which is
\begin{align}
    \int (\alpha,\beta) = \int \alpha \wedge \star \beta\ ,
\end{align}
for any two differential forms $\alpha,\beta$.
If the two forms are not of the same degree, this inner product is automatically zero. Thus, the inner product generalizes directly to multiforms for which it matches components of the same degree. For the rest of this discussion, $\alpha, \beta$ denote multiforms.

Under the integral, it holds that
\begin{align}
    \int \alpha \wedge \star \beta = \int \star(\alpha \vee \beta)\ ,
\end{align}
by uniqueness of the Hodge star isomorphism, which maps only the fully contracted component of $\alpha\vee\beta$ to a top-form.
Due to the associativity of the Clifford product, it is
\begin{align}
    \int (\alpha,\beta\vee\gamma) = \int \star(\alpha \vee \beta \vee \gamma) = \int (\alpha \vee \beta , \gamma)\ .
\end{align}
Together with the symmetry of the Hodge inner product, this implies the cyclicity
\begin{align}
    \int (\alpha,\beta\vee \gamma) = \int (\gamma,\alpha\vee \beta) = \int (\beta, \gamma\vee\alpha)\ .
\end{align}

The second property of the inner product is that we can integrate the Kähler-Dirac operator by parts, as it is
\begin{align}
    &\int (\alpha, \mathrm d \beta) = \int (\mathrm d^\dagger \alpha, \beta) \nonumber\\
    \Rightarrow &\int (\alpha, K\beta) = \int (\alpha, (\mathrm d + \mathrm d^\dagger )\beta)\nonumber \\= &\int ((\mathrm d^\dagger + \mathrm d) \alpha, \beta) = \int (K\alpha,\beta)\ ,
\end{align}
up to boundary terms.

\bibliography{bibYM}

\end{document}